\documentclass{iopart}
\usepackage{graphicx,amssymb} 

\begin{document}

\title{Topological quantization of ensemble averages}

\author{Emil Prodan}

\address{Department of Physics, Yeshiva University, New York, NY 10016}

\begin{abstract}
We define the current of a quantum observable and, under well defined conditions, we connect its ensemble average to the index of a Fredholm operator. The present work builds on a formalism developed by  Kellendonk and Schulz-Baldes \cite{Kellendonk:2004p597} to study the quantization of edge currents for continuous magnetic Schroedinger operators. The generalization given here may be a useful tool to scientists looking for novel manifestations of the topological quantization. As a new application, we show that the differential conductance of atomic wires is given by the index of a certain operator. We also comment on how the formalism can be used to probe the existence of edge states.
\end{abstract}

\maketitle

\section{Introduction} 

Observables in Quantum Mechanics are represented by self-adjoint operators. A large class of self-adjoint operators have discrete spectrum, which means that some  observables will take only a set of quantized values. This kind of quantization is now well understood. What is more difficult to understand is the quantization of macroscopic variables, quantization that takes place even at finite temperatures or at non-equilibrium, when the experimentally measured values are actually ensembles averages. There are a number of quite spectacular examples of this kind, like the quantization of conductance in metallic atomic wires \cite{Terabe:2005cr},  quantization of Hall conductance \cite{Klitzing:1980dq}, quantization of surface forces \cite{Kellendonk:2004p595} and so on. Without any exception, the phenomenon of quantization was observed to be robust under continuous deformations of the system. This automatically imply that some topological properties of the system are responsible for this phenomenon. 

The topological aspects of bulk integer quantum Hall effect were understood for quite some time \cite{Richter:1999p731, Avron:2003p732}. However, the quantization of the edge currents in the presence of random scatterers was rigorously solved only recently \cite{SchulzBaldes:2000p599, Kellendonk:2002p598, Kellendonk:2004p597, Elbau:2002qf, Elgart:2005rc}. The solution given in Ref.~\cite{Kellendonk:2004p597} for continuum magnetic Shcroedinger opperators connects the problem of ensemble average of edge currents to powerful ideas in non-commutative geometry.  Actually, this solution provides a platform that opened the possibility of many additional applications and further generalizations \cite{Kellendonk:nx}.

In this article, we formulate an abstract statement about the topological quantization of the current of a general observable. We were able to formulate a set of general conditions that are sufficient for the observation of the topological quantization. The backbone of the proof of our main statement follows Ref.~\cite{Kellendonk:2004p597}, but certain parts of the proof required different approaches, manly because we were lacking certain direct estimates related to the concrete system considered in Ref.~\cite{Kellendonk:2004p597}. For example, the last step in our proof is substantially different from that of Ref.~\cite{Kellendonk:2004p597} and involves a non-commutative residue theorem that is formulated and demonstrated in the present paper. We also want to add that the set of general conditions formulated here can be verified, in concrete examples, using alternative paths from those of Ref.~\cite{Kellendonk:2004p597}. This is illustrated in Ref.~\cite{Prodan:2008ai} where we studied the quantization of edge currents in Chern insulators.

The generalization presented in this paper was motivated by our search for manifestations of the topological quantization in new systems (the Spin-Hall system was one of them). Our quest helped us put the ideas into a general framework that now states that,  under a well defined set of conditions, the ensemble average of the current of a general observable, suitably defined, is equal to the index of a Fredholm operator. It remains to be seen if an entirely new observable can be identified and what are the physical conditions in which the quantization of its current takes place (the mathematical conditions are formulated in this paper). Following, to some extent, the already well established examples, we found a new application of the formalism, namely, to the quantization of the differential conductance in metallic atomic. This application is presented at the end of the paper. 

We also sketch at the end of the paper how the result can be used to study the existence of edge states, in particular, we argue that such an approach will be able to handle cases that cannot be treated with the existing formalisms.

\section{The problem and the result.}

Consider the following set of assumptions (called ${\cal A}$): \medskip

\begin{itemize}

\item The existence of a separable Hilbert space $\cal{H}$.

\item The existence of a family of self-adjoint Hamiltonians $H_\omega$ on $\cal{H}$. The index $\omega$ lives in the space $\Omega$. 

\item The existence of an observable $X$ with continuous spectrum that extends from $-\infty$ to $+\infty$, and such that $[X,H_\omega]$ is $H_\omega$-relative bounded. The bound is assumed uniform in $\omega$.

\item The existence of a 1-parameter, strongly continuous unitary group $u_a$ such that:
\begin{equation}
u_a X u_a^{-1}=X+a.
\end{equation}

\item The covariance of the family $H_\omega$ relative to the unitary transformations $u_a$, i.e. the existence of a flow $t_a$ on $\Omega$ such that:
\begin{equation}
u_a H_\omega u_a^{-1} = H_{t_a \omega}, \ \mbox{for all} \ \omega \in \Omega. \medskip
\end{equation}

\item The ergodicity of the flow $t_a$ over $\Omega$.
 
\item The existence of a probability measure $dP(\omega)$ over the compact set $\Omega$, invariant to the flow $t_a$.

\end{itemize}

The above statements describe the physical systems we focus to in this paper. Below we state the key properties these systems must have.  We are concerned with certain parts of the energy spectrum, where the Hamiltonians $H_\omega$ have a special behavior. Assume that such a spectral interval $s$ was identified and let $F(x)$ be an arbitrarily shaped, smooth function that is equal to 1 below $s$ and to 0  above $s$.  Using the spectral calculus, we define the following unitary operators:
\begin{equation}\label{uomega}
U_\omega = e^{-2 \pi i F(H_\omega)}.
\end{equation}
The special behavior inside the spectral interval $s$ is related to the following two properties (called ${\cal P}$):\medskip

\begin{itemize}
\item Let $\pi_\Delta (x)$ be the spectral projector of $X$ onto the interval $[x,x+\Delta)$. We require that $(U_\omega-I) \pi_\Delta (x)$ is Hilbert-Schmidt for any finite $\Delta$ and any $x\in {\bf R}$. We will use the notation:
\begin{equation}
K_\Delta (x,x';\omega)= \| \pi_\Delta (x) (U_\omega-I)\pi_\Delta (x') \| ^2 _{\mbox{\tiny{HS}}}.
\end{equation}

\item The following upper bound
\begin{equation}\label{decay}
K_\Delta (x,x';\omega) \leq \frac{c_\Delta}{1+|x-x'|^\alpha}
\end{equation}
holds true, with $\alpha > 3$ and $c_\Delta$ behaves as $\Delta^\beta$ with $\beta>1$ in the limit $\Delta$ goes to zero. \medskip
\end{itemize}
We add a few notes about the above points. If $(U_\omega-I)$ has a kernel that is continuous, then $c_\Delta$ behaves as $\Delta^2$ in the limit $\Delta\rightarrow 0$. This means, the condition we require is weaker than the continuity of the kernel. $c_\Delta$ will depend on the shape of $F(x)$, but it is almost surely independent of $\omega$. This follows from the fact that $K_\Delta (x,x';t_a \omega)=K_\Delta (x+a,x'+a;\omega)$ and because $t_a$ acts ergodically on $\Omega$. Thus it will not be restrictive to assume that $c_\Delta$ is independent of $\omega$.

We now define the trace (notation $\mbox{tr}_0$) over the "states of zero expectation" value for $X$:
\begin{equation}
\mbox{tr}_0 \{A\}=\lim_{\Delta \rightarrow 0} \frac{1}{\Delta} \mbox{Tr} \{\pi_\Delta (0)A\pi_\Delta(0)\}.
\end{equation}
The domain of this operation is described in Proposition 1. In the calculations that follows, tr${_0}$ is always finite due to the assumptions ${\cal P}$. We use $\mbox{tr}_0$ to define the current of the observable $X$. The quantum time evolution of $X$ is given by $X_\omega(t)=e^{itH_\omega}X e^{-itH_\omega}$ and its time derivative satisfies the identity: $\mbox{d}X_\omega(t)/\mbox{d}t=i[H_\omega,X_\omega(t)]$. We define the current of $X$ as the expectation value of its time derivative evaluated at the present time (chose here to be t=0), with the expectation value taken only over the states of zero $X$:
\begin{equation}
J_\omega = \left . \mbox{tr}_0 \left\{\rho(H_\omega) \frac{\mbox{d}X_\omega(t)}{\mbox{d}t}\right\} \right |_{t=0}=i\left . \mbox{tr}_0 \left\{\rho(H_\omega) [H_\omega,X] \right\} \right.
\end{equation}
Here $\rho(H_\omega)$ is the statistical distribution of the quantum states. We assume $\rho(x)$ to be smooth, with support in the spectral interval $s$ and normalized as $\int \rho(x)dx$=1. We define $F(x)=\int_{x}^\infty$ (or $\rho(x)=-\mbox{d}F(x)/\mbox{d}x$), which is a smooth function equal to 1 below $s$ and to 0 above $s$.

The quantity $\mbox{tr}_0 \{ \rho(H_\omega) \mbox{d}X / \mbox{d}t \}$ is a natural definition of the current of $X$. For example, if $X$ is the spatial coordinate $x$, then $\mbox{d}x/\mbox{d}t$ is the observable corresponding to the current density of the particles in the $x$ direction. To get the current from the current density, we have to integrate the later over a section of the space normal to the $x$ direction. The $\mbox{tr}_0$ does just that.\medskip

\noindent {\bf Main Theorem.}  Assume a system described by ${\cal A}$. Assume that the properties ${\cal P}$ hold true on a spectral interval $s$. If $\pi_+$ is the projector onto the positive spectrum of $X$, then $\pi_+ U_\omega \pi_+$ is in the Fredholm class and
\begin{equation}\label{main}
\int_\Omega dP(\omega) \ J_\omega  = \frac{1}{2\pi}   \mbox{Ind} \left \{\pi_+ U_\omega \pi_+ \right \}.
\end{equation} 
The index is independent of the particular choice of $\rho(x)$ and is almost surely independent of $\omega$.\medskip

\section{Technical results}

Before starting the proof of our main statement, we want to point out two important consequences of the properties ${\cal P}$ introduced above. Along this paper, the following notations $\| \ \|$, $\| \ \|_{\mbox{\tiny{HS}}}$ and  $\| \ \|_1$ represent the operator, Hilbert-Schmidt and trace norms, respectively. Also, $\pi_\pm$ represent the spectral projectors of $X$ onto the positive/negative spectrum and $\Sigma\equiv \pi_+ - \pi_-$.

It is convenient to introduce from the beginning an approximate spectral projector onto the support of $\mbox{d}F/\mbox{d}x$. For this, we consider a smooth function $G(x)$ with same properties as $F$ plus the property that is equal to $1/2$ on the support of $\mbox{d}F/\mbox{d}x$. In this case,
\begin{equation}
\pi_s = \frac{1}{2}\left (I-e^{-2\pi i G(H_\omega)}\right)
\end{equation}
leaves invariant the states inside the spectral interval of support of $\mbox{d}F/\mbox{d}x$. By construction, $\pi_s$ has similar properties as $U_\omega - I$, which are stated below.

\medskip

\noindent  {\bf Proposion 1.} Let $\chi(x)$ be  a smooth, compactly supported functions. The following statements are true:
\begin{enumerate}
\item $(U_\omega-I)\chi(X)$ are Hilbert-Schmidt.  \\
\item $[\Sigma, U_\omega]$ are Hilbert-Schmidt. \\
\item $[X,U_\omega]\chi(X)$ are Hilbert-Schmidt.
\end{enumerate}
\noindent The Hilbert-Schimdt norms of the above operators are uniformly bounded relative to $\omega$.\medskip

\noindent {\it Proof.} i) We divide the real axis in equal intervals as shown in Fig.~\ref{division}. We denote the spectral projector of $X$ onto the interval $\Delta_n=[x_n,x_n+\Delta)$ by $\pi_n$. We have:
\begin{equation}
\begin{array}{c}
\| (U_\omega-I)\chi(X)\|_{\mbox{\tiny{HS}}}^2 = \sum\limits_{n,n'} \| \pi_{n'} (U_\omega-I)\chi(X)\pi_{n} \|_{\mbox{\tiny{HS}}}^2 \medskip \\
\leq \sum\limits_{n,n'} \| \pi_{n'} (U_\omega-I)\pi_{n} \|_{\mbox{\tiny{HS}}}^2 \|\chi(X)\pi_{n}\|^2 \medskip \\
= \sum\limits_{n,n' } K_\Delta (x_{n'},x_{n})\sup\limits_{y_n\in \Delta_n} \chi(y_n)^2.
\end{array}
\end{equation}
The final sum converges to a finite value due to our requirement in ${\cal P}$ that $\alpha>3$ and the compactness of the support of $\chi$.\smallskip

\noindent ii) We proceed as follows.
\begin{equation}
\begin{array}{c}
\| [\Sigma,U_\omega] \|_{\mbox{\tiny{HS}}}^2 = \sum\limits_{n,n'} \| \pi_n [\Sigma,U_\omega]\pi_{n'} \|_{\mbox{\tiny{HS}}}^2 \medskip \\
=4\sum\limits_{n\cdot n' \leq 0} \| \pi_n (U_\omega-I)\pi_{n'} \|_{\mbox{\tiny{HS}}}^2 = 4\sum\limits_{n\cdot n' \leq 0} K_\Delta (x_n,x_{n'}).
\end{array}
\end{equation}
In the last two sums, we must exclude the term $n$=$n'$=0. The final sum converges to a finite value due to our requirement in ${\cal P}$ that $\alpha>3$.\smallskip

\noindent iii) We use the following equivalent expression for the commutator $[X,U_\omega]$:
\begin{equation}
\begin{array}{c}
[X,U_\omega]=\sum\limits_{n,n'}  (x_n-x_{n'}) \pi_n (U_\omega-I) \pi_{n'} \medskip \\
+\sum\limits_{n,n'}\{(X-x_n)\pi_n (U_\omega-I) \pi_{n'}+\pi_n (U_\omega-I) \pi_{n'}(x_{n'}-X)\}.
\end{array}
\end{equation}
to obtain:
\begin{equation}
\begin{array}{c}
\| [X,U_\omega]\chi(X) \|_{\mbox{\tiny{HS}}}^2  \medskip \\
\leq \sum\limits_{n,n'}  (|x_n-x_{n'}|+\|(X-x_n)\pi_n\|+\|(X-x_{n'})\pi_{n'}\|)^2  \medskip \\
  \times \| \pi_n (U_\omega-I) \pi_{n'} \|_{\mbox{\tiny{HS}}}^2 \ \|\chi(x_{n'})\|^2 \nonumber \medskip \\
 \leq  \sum\limits_{n,n'}  \frac{c_\Delta (|x_n-x_{n'}|+2 \Delta)^2\|\chi(x_{n'})\|^2 }{1+|x_n-x_{n'}|^\alpha} . \nonumber
\end{array}
\end{equation} 
The sum convergences to a finite value due to our requirement in ${\cal P}$ that $\alpha > 3$.

\section{Basic properties of tr$_0$}

Here we list three properties of tr$_0$, essential for the proof of our main statement.\medskip

\noindent{\bf Property 1.} Let $\{A_\omega\}_{\omega \in \Omega}$ be a covariant family of operators such that $\| h(X)A_\omega h(X)\|_1$ $< t$, for any smooth $h(x)$ of compact support. For fixed $h$, the upper bound $t$ is assumed to be almost surely independent of $\omega$. Then, if $\chi(x)$ is a smooth, compactly supported function such that $\int \chi(x)^2 $=1, then
\begin{equation}
\int dP(\omega)\mbox{tr}_0 \{A_\omega \} =\int dP(\omega) \mbox{Tr}\{\chi(X)  A_\omega \chi(X)  \} <t.\medskip
\end{equation}
We can conclude from this statement that: $\mbox{tr}_0 \{A_\omega \}$ is almost surely finite if  $\| h(X)A_\omega h(X)\|_1$ $< t<\infty$ for all $\omega\in \Omega$.\smallskip

\noindent{\bf Property 2.}  Let $\{A_\omega\}_{\omega \in \Omega}$ and $\{B_\omega\}_{\omega\in \Omega}$ be two covariant families of operators such that $\chi(X)A_\omega$, $A_\omega\chi(X)$, $\chi(X)B_\omega$ and $B_\omega\chi(X)$ are Hilbert-Schmidt. For fixed $\chi(x)$,  the Hilbert-Schmidt norms  of these operators are assumed uniformly bounded in $\omega$. Then:
\begin{equation}
\int dP(\omega) \mbox{tr}_0 \{A_\omega B_\omega \}=\int dP(\omega) \mbox{tr}_0 \{B_\omega A_\omega \}<\infty.\medskip
\end{equation}

\noindent{\bf Property 3.} Let $\{A_\omega\}_{\omega \in \Omega}$ be a covariant family of operators such that $\| \chi(X)A_\omega \chi(X)\|_1$$<t$, for all $\omega$ except a possible zero measure subset of $\Omega$. Then
\begin{equation}
 \int dP(\omega) \mbox{tr}_0 \{[X,A_\omega] \}=0.\medskip
\end{equation}
\medskip

\noindent{\it Proof of Property 1.} We use the invariance of the trace under unitary transformations to write first:
\begin{equation}
\begin{array}{c}
 \frac{1}{\Delta}\mbox{Tr}\{\pi_0 A_\omega \pi_0 \}
  = \sum\limits_n\chi(y_n)^2 \mbox{Tr}\{u_{x_n} \pi_0 A_\omega  \pi_0u_{x_n}^*\} \medskip \\
  =\sum\limits_n\chi(y_n)^2 \mbox{Tr}\{ \pi_n A_{t_{x_n}\omega}  \pi_n\},
  \end{array}
\end{equation}
where $y_n$$\in$$\Delta_n$ were chosen such that $\Delta \sum\limits_n \chi(y_n)^2$=1. Note that this is always possible since $\chi(x)$ is smooth and $\int \chi(x)^2$=1.
Then, using the invariance of the measure $P(\omega)$, we have
\begin{equation}
\int dP(\omega) \frac{1}{\Delta}\mbox{Tr}\{\pi_0 A_\omega \pi_0 \}=\int dP(\omega)  \mbox{Tr}\{\chi_\Delta(X) A_\omega \chi_\Delta(X) \}, 
\end{equation}
where $\chi_\Delta(X) = \sum\limits_n \chi(y_n)\pi_n$. We write
\begin{equation}
\begin{array}{c}
\int dP(\omega) \frac{1}{\Delta}\mbox{Tr}\{\pi_0 A_\omega \pi_0 \}=\int dP(\omega) [\mbox{Tr}\{\chi(X) A_\omega \chi (X) \} \medskip \\
+\mbox{Tr}\{\chi_\Delta(X) A_\omega \chi_\Delta(X)-\chi(X) A_\omega \chi(X) \}],
\end{array} 
\end{equation}
and we show that the last term inside the integral goes to zero as $\Delta$ goes to zero. Indeed, if $I$ is an interval large enough so it contains the supports of $\chi(X)$ and $\chi_\Delta(X)$ (for any small $\Delta$) and $h_I$ is a smooth, compactly supported function that is equal to 1 inside the interval $I$, then
\begin{equation}
\begin{array}{c}
|\mbox{Tr}\{\chi_\Delta(X) A_\omega \chi_\Delta(X) -\chi(X) A_\omega \chi(X) \}| \medskip\\
=|\mbox{Tr}\{[\chi_\Delta(X)-\chi(X)] A_\omega \chi_\Delta(X) + \chi(X) A_\omega (\chi_\Delta(X)-\chi(X)]| \}\medskip \\
 =|\mbox{Tr}\{[\chi_\Delta(X)-\chi(X)] h_I(X) A_\omega h_I(X)\chi_\Delta(X) \medskip \\
 + \chi(X) h_I(X) A_\omega h_I(X) (\chi_\Delta(X)-\chi(X)]| \}\medskip \\
 \leq \|\chi_\Delta(X)-\chi(X)\| (\|\chi_\Delta(X)\|+\|\chi(X)\|)\|h_I(X) A_\omega h_I(X)\|_1, 
\end{array}
\end{equation}
and the first term of the last row converges to zero as $\Delta$ goes to zero since $\chi(x)$ is smooth.\medskip

\noindent{\it Proof of Property 2.} Let $\pi (M)=\sum\limits_{-M}^M \pi_n$. Then $\pi_0 A_\omega  \pi(M) B_\omega \pi_0$ is trace class and
\begin{equation}
\begin{array}{c}
\int dP(\omega) \mbox{Tr}\{\pi_0 A_\omega  \pi(M) B_\omega \pi_0 \} \medskip \\
=\int dP(\omega) \sum\limits_{n=-M}^M\mbox{Tr}\{\pi_0 A_\omega  \pi_n B_\omega \pi_0 \} \medskip \\
=\int dP(\omega) \sum\limits_{n=-M}^M\mbox{Tr}\{\pi_n B_\omega  \pi_0 A_\omega \pi_n \} 
\end{array}
\end{equation}
At this point we use the invariance of the trace (on trace class operators) under the unitary transformations to continue:
\begin{equation}
\begin{array}{c}
\ldots =\int dP(\omega) \sum\limits_{n=-M}^M \mbox{Tr}\{u_{x_n}\pi_n B_\omega  \pi_0 A_\omega \pi_n u_{x_n}^*\} \medskip \\
=\int dP(\omega) \sum\limits_{n=-M}^M \mbox{Tr}\{\pi_{0} B_{t_{x_n}\omega}  \pi_n A_{t_{x_n}\omega} \pi_{0}\} \medskip \\
=\int dP(\omega) \sum\limits_{n=-M}^M \mbox{Tr} \{\pi_0 B_{\omega} \pi_{n} A_{\omega}  \pi_0\}.
\end{array}
\end{equation}
At the end of above argument we used the invariance of $P(\omega)$ relative to the flow $t_a$. We then have that
\begin{equation}
\int dP(\omega) \mbox{Tr}\{\pi_0 A_\omega  \pi(M) B_\omega \pi_0 \}=\int dP(\omega) \mbox{Tr}\{\pi_0 B_\omega  \pi(M) A_\omega \pi_0 \},
\end{equation}
and the affirmation follows by letting $M$ go to infinity and then $\Delta$ to zero.\medskip

\noindent{\it Proof of Property 3.} Since $[X,A_\omega]$ is also a covariant family and $\chi(X)[X,A_\omega]\chi(X)$ is trace class, we have
\begin{equation}
 \int dP(\omega) \mbox{tr}_0 \{[X,A_\omega] \}=\int dP(\omega) \mbox{Tr}\{\chi(X)  [X,A_\omega] \chi(X)  \}.
\end{equation}
The affirmation follows from the fact that if $Q$ is trace class and $R$ is bounded, then $\mbox{Tr}\{QR\}= \mbox{Tr}\{RQ\}$. This property is proven, for example, in Ref.~\cite{Simon:2005p998}.

\section{Proof of the main statement}

We have the following identities:
\begin{equation}
\begin{array}{c}
\pi_+ -\pi_+ U_\omega^* \pi_+ U_\omega \pi_+=-\frac{1}{4}\pi_+ [\Sigma,U_\omega^*][\Sigma,U_\omega] \medskip \\
\pi_+ -\pi_+ U_\omega \pi_+ U_\omega^* \pi_+=-\frac{1}{4}\pi_+ [\Sigma,U_\omega^][\Sigma,U_\omega^*],
\end{array}
\end{equation} 
which, together with Proposition 1, imply that $\pi_+ -\pi_+ U_\omega^* \pi_+ U_\omega \pi_+$ and $\pi_+ -\pi_+ U_\omega \pi_+ U_\omega^* \pi_+$ are trace class. Given this, the Fedosov principle \cite{Kellendonk:2002p598} tells us that $\pi_+ U_\omega \pi_+$ is in the Fredholm class and that:
\begin{equation}
\begin{array}{c}
\mbox{Ind} \{\pi_+ U_\omega \pi_+\} \medskip \\
 =-\frac{1}{4} \mbox{Tr} \{\pi_+ [\Sigma,U_\omega^*][\Sigma,U_\omega] \} +  \frac{1}{4}\mbox{Tr} \{\pi_+ [\Sigma,U_\omega][\Sigma,U_\omega^*]\},
 \end{array}
 \end{equation}
which can be brought to the following compact form:
\begin{equation}
\mbox{Index}\{\pi_+ U_\omega \pi_+\}=-\frac{1}{4}\mbox{Tr}\{\Sigma[\Sigma,U_\omega^*][\Sigma,U_\omega]\},
\end{equation}
We reformulate it a slightly different format using the following result.\medskip

\noindent {\bf Lemma 2.} Consider two bounded operators $A$ and $B$ such that $[\Sigma, A]$ and $[\Sigma,B]$ are Hilbert-Schmidt (see Proposition 1). Then:
\begin{equation}
\mbox{Tr}\{\Sigma[\Sigma,A][\Sigma,B]\}=2\sum\limits_{\beta=\pm} \mbox{Tr}\{\pi_\beta A[\Sigma,B]\pi_\beta\}.
\end{equation}
{\it Proof.} Indeed, since $[\Sigma,A]$ is Hilbert-Schmidt, $\pi_\beta A \pi_{-\beta}$ is Hilbert-Schmidt. Then the operator $\pi_\beta A[\Sigma,B]\pi_\beta$ is trace class, since
\begin{equation}\label{inter1}
\pi_\beta A[\Sigma,B]\pi_\beta=\pi_\beta A\pi_{-\beta}[\Sigma,B]\pi_\beta.
\end{equation}
Moreover,
\begin{equation}
\begin{array}{c}
\mbox{Tr}\{\Sigma[\Sigma,A][\Sigma,B]\}=\sum\limits_\beta \mbox{Tr}\{\pi_\beta \Sigma[\Sigma,A][\Sigma,B]\pi_\beta \} \medskip \\
=\sum\limits_\beta \mbox{Tr}\{\pi_\beta \Sigma[\Sigma,A]\pi_{-\beta}[\Sigma,B]\pi_\beta \} \medskip \\
=2\sum\limits_\beta \mbox{Tr}\{\pi_\beta A\pi_{-\beta}[\Sigma,B]\pi_\beta\},
\end{array}
\end{equation}
and the statement follows from Eq.~\ref{inter1}.\medskip

\begin{figure}
 \center
 \includegraphics[width=6cm]{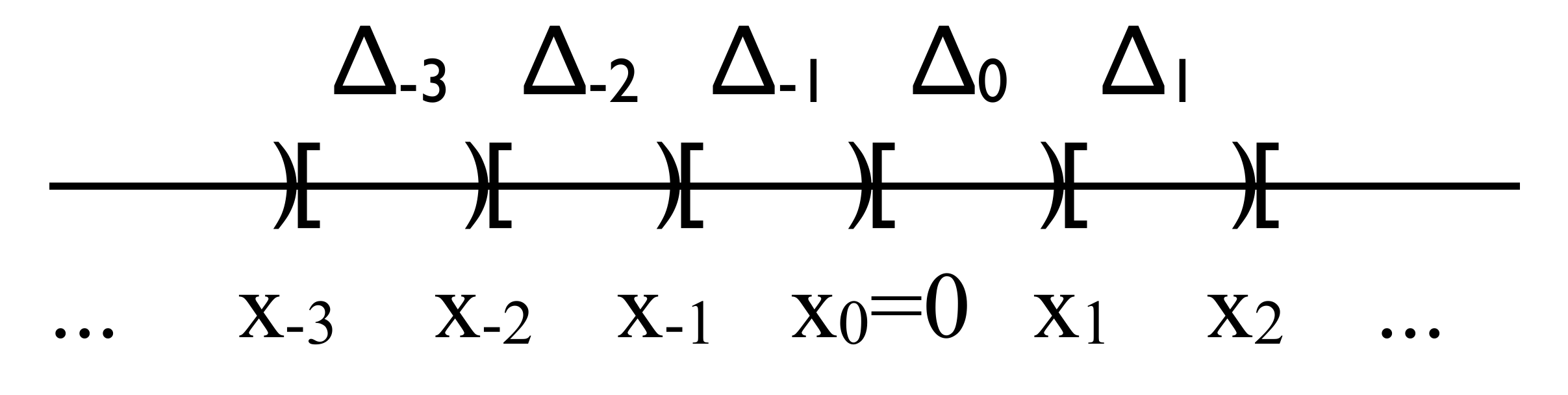}
 \caption{Division of the real axis in equal intervals.}.
 \label{division}
\end{figure}

We continue the main proof. At this point we can write:
\begin{equation}
\begin{array}{c}\label{ind}
\mbox{Ind}\{\pi_+ U_\omega \pi_+\}=-\frac{1}{2}\sum\limits_\beta \mbox{Tr}\{\pi_\beta (U_\omega^*-I)[\Sigma,U_\omega]\pi_\beta\}   \medskip \\
=-\frac{1}{2}\sum\limits_{n} \mbox{Tr}\{\pi_n (U_\omega^*-I)[\Sigma,U_\omega]\pi_n\} .
\end{array}
\end{equation}
We use this formula to prove the invariance of the index. Let $a$ be an arbitrary positive number. The case when $a$ is negative can be handled in a similar way. We will compare the index of $\pi_+U_\omega \pi_+$ and $\pi_+ U_{-t_a \omega} \pi_+$, which are both in the Fredholm class. For this, we take the division of the real axis such that $x_1=a$, in which case we have:
\begin{equation}
\begin{array}{c}
\mbox{Ind}\{\pi_+U_\omega \pi_+\}-\mbox{Ind}\{\pi_+ U_{-t_a \omega} \pi_+\}\medskip \\
=\mbox{Ind}\{\pi_+U_\omega \pi_+\}-\mbox{Ind}\{u_{-a}\pi_+ u_{-a}^*U_{\omega} u_{-a}\pi_+ u_{-a}^*\} \medskip \\
=-\frac{1}{2}\sum\limits_{n} \mbox{Tr}\{\pi_n (U_\omega^*-I)[\Sigma-u_{-a} \Sigma u_{-a}^*,U_\omega]\pi_n\} \medskip \\
=-\sum\limits_{n} \mbox{Tr}\{\pi_n (U_\omega^*-I)[\pi_0,U_\omega]\pi_n\}. \medskip \\
\end{array}
\end{equation}
Since $(U_\omega -I)\pi_0$ is Hilbert-Schmidt, we can open the commutator and continue:
\begin{equation}
\begin{array}{c}
\ldots =-\sum\limits_{n} \mbox{Tr}\{\pi_n (U_\omega^*-I)\pi_0(U_\omega-I)\pi_n\} \medskip \\
+ \mbox{Tr}\{\pi_0 (U_\omega^*-I)(U_\omega-I)\pi_0 \} \medskip \\
=-\mbox{Tr}\{(U_\omega^*-I)\pi_0(U_\omega-I)-\pi_0 (U_\omega-I)(U_\omega^*-I)\pi_0 \} \medskip \\
\end{array}
\end{equation}
which is identically zero since $\mbox{Tr}\{A B\}$=$\mbox{Tr}\{BA\}$ for $A$ and $B$ Hilbert-Schmidt operators. Thus, we showed that the index is invariant to the flow of $t_a$, which acts ergodically on $\Omega$. Consequently, $\mbox{Ind} \{\pi_+ U_\omega \pi^+\} $ is constant for all $\omega \in \Omega$, excepting a possible zero measure subset. 

\begin{figure}
\center
 \includegraphics[width=5cm]{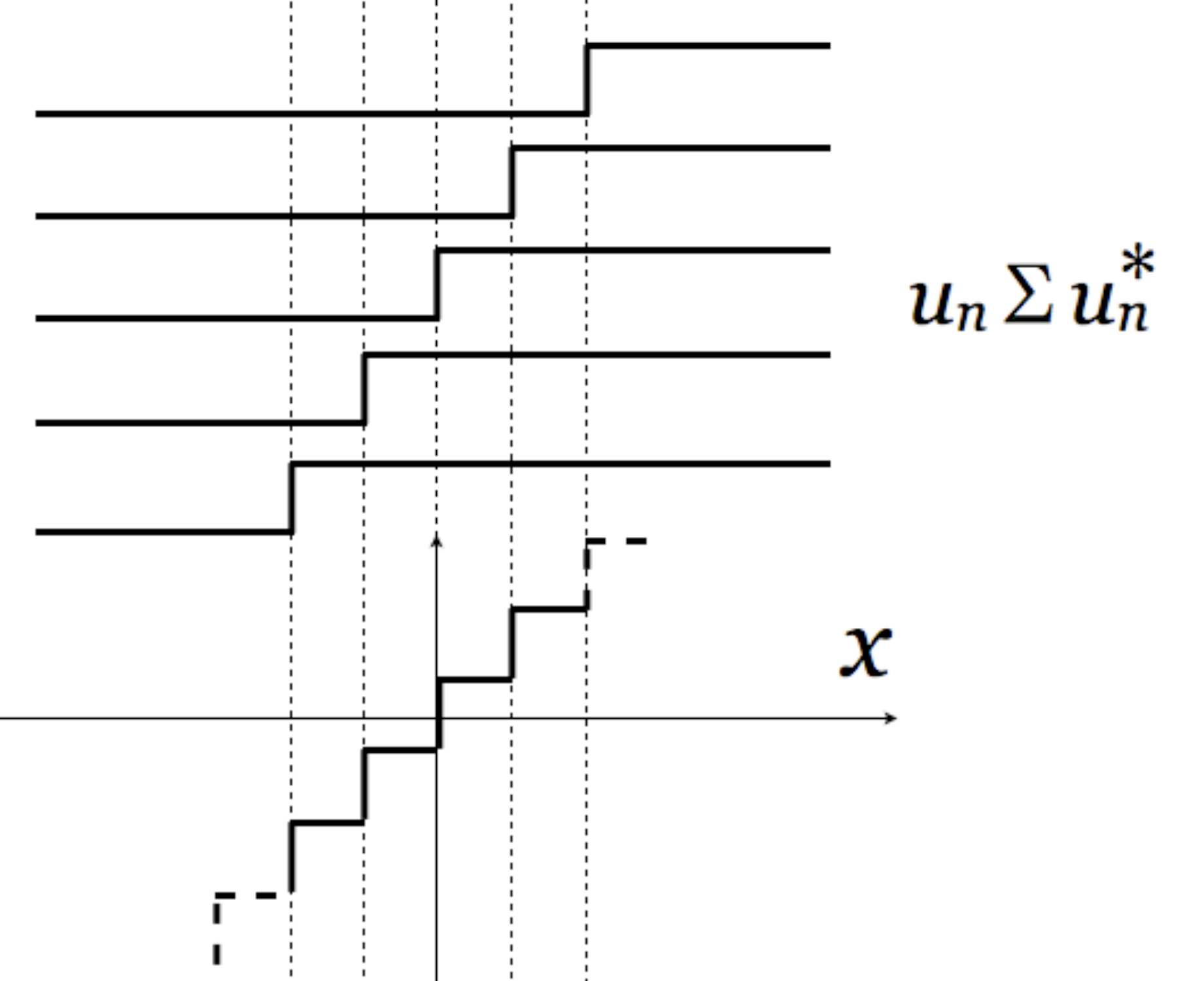}\\
 \caption{A graphical representation of $\sum u_{x_n}\Sigma u_{x_n}^*=\sum (2n+1)\pi_n$. The top lines represent the spectral representations of $u_{x_n}\Sigma u_{x_n}$, which are sign functions shifted by $x_n$. The sum of the top lines results in the stair like function shown by the bottom line.}  
 \label{fig2}
 \end{figure}

We continue from Eq.~\ref{ind} and consider the average over $\omega$. This operation has no effect on the left hand side. On the right hand side, using the fact that the trace of trace-class operators is invariant to unitary transformations and that the measure $dP(\omega)$ is invariant to the flow $t_a$, we write:
\begin{equation}\label{step1}
\begin{array}{c}
\mbox{Ind}\{\pi_+ U_\omega \pi_+\}  \medskip \\
 =-\frac{1}{2}\int dP(\omega)\sum\limits_{n}  \mbox{Tr}\{u_{x_n}\pi_n (U_\omega^*-I)[\Sigma,U_\omega]\pi_n u_{x_n}^* \} \medskip \\
=-\frac{1}{2}\int dP(\omega)\sum\limits_{n} \mbox{Tr}\{\pi_0 (U_{t_{x_n}\omega}^*-I)[u_{x_n}\Sigma u_{x_n}^*,U_{t_{x_n}\omega}]\pi_0 \} \medskip \\
=-\int dP(\omega) \frac{1}{\Delta}\mbox{Tr}\{\pi_0 (U_\omega^*-I)[\frac{\Delta}{2} \sum\limits_{n}u_{x_n}\Sigma u_{x_n}^*,U_\omega]\pi_0 \}.
\end{array}
\end{equation}
We now use the key observation that:
\begin{equation}
\frac{\Delta}{2}\sum\limits_{n}u_{x_n}\Sigma u_{x_n}^*=\sum\limits_{n}(n+1/2)\Delta \pi_n\equiv X_\Delta.
\end{equation}
A graphical representation of this relation is given in Fig.~\ref{fig2}. Now
\begin{equation}
\begin{array}{c}
\frac{1}{\Delta}|\mbox{Tr}\{\pi_0 (U_\omega^*-I)[X_\Delta-X,U_\omega]\pi_0 \}| \medskip \\
=\frac{1}{\Delta}|\sum\limits_n\mbox{Tr}\{\pi_0 (U_\omega^*-I)\pi_n [X_\Delta-X,U_\omega]\pi_0 \}| \medskip \\
\leq \frac{2}{\Delta}\sum\limits_n\| \pi_0 (U_\omega^*-I)\pi_n\|_{\mbox{\tiny{HS}}} \|\pi_n (U_\omega-I)\pi_0 \}\|_{\mbox{\tiny{HS}}} \|X_\Delta - X\| \medskip \\
=\sum\limits_n\ K_\Delta(x_n,x_0) \leq \frac{c_\Delta}{\Delta} \sum\limits_n \frac{\Delta}{1+|x_n-x_0|^\alpha}
\end{array}
\end{equation}
In the limit $\Delta$$\rightarrow$0, the sum, above, converges uniformly to $\int 1/(1+|x|^\alpha)dx$, which is finite due to our requirement that $\alpha>3$, while the term in front of the sum goes to zero due to our requirement that $c_\Delta$ behave as $\Delta^\beta$, with $\beta >1$.  This allows us to replace $X_\Delta$ by $X$ in Eq.~\ref{step1} and to arrive at the intermediated conclusion that:
\begin{equation}\label{last}
 \mbox{Ind}\{\pi _+U_\omega \pi_+\}=-\int dP(\omega) \ \mbox{tr}_0\{ (U_{\omega}^*-I)[X, U_{\omega}] \}.
\end{equation}
The integrand of the last integral is finite, fact that can be seen from the second point of Proposition 1 and Property 1. To advance with our proof, we need the following result.\medskip

\noindent{\bf A Non-Commutative Residue Theorem.} Let $f(z)$ be analytic in a strip around the unit circle. If $\{U_\omega\}_{\omega \in \Omega}$ a covariant family of unitary operators such that $\chi(X)(U_\omega-I)$ and $[X, U_\omega]\chi(X)$ are Hilbert-Schmidt, then:
\begin{equation}
\begin{array}{c}
\int dP(\omega) \mbox{tr}_0\{(f(U_\omega)-f(I))[X, U_\omega]\} \medskip \\
=b_1 \int dP(\omega)\mbox{tr}_0 \{(U_\omega^*-I) [X,U_\omega]\},
\end{array}
\end{equation}
where $b_1$ is the coefficient appearing in the Laurent expansion:
\begin{equation}
f(z)=\sum\limits_{n=1,\infty} b_n z^{-n} + \sum\limits_{n=0,\infty} a_n z^n.
\end{equation}\medskip

We continue from Eq.~\ref{last} where,  using the Non-Commutative Residue Theorem, we replace $U^*_\omega - I$ by $\frac{1}{b_1}(f(U_\omega)-f(I))$, with $f(z)$ analytic in a strip around the unit circle. This point will become crucial at a later point in the proof. Also, to ensure that everything stays finite and convergent we proceed as follows:
\begin{equation}\label{ref1}
\begin{array}{c}
 \mbox{Ind}\{\pi _+U_\omega \pi_+\} \medskip \\
=-\frac{1}{b_1}\int dP(\omega) \ \mbox{tr}_0\{ (f(U_{\omega})-f(I))[X, U_{\omega}] \} \medskip \\
=-\frac{1}{b_1}\int dP(\omega) \ \mbox{tr}_0\{\pi_s^2 (f(U_{\omega})-f(I))[X, U_{\omega}] \}. 
\end{array}
\end{equation}
and we use Property 2 to move $\pi_s^2$ all the way to the right, inside $\mbox{tr}_0$. We evaluate the commutator using Duhamel's identity
\begin{equation}
\begin{array}{c}
[X, U_{\omega}]=-\int dt \ \tilde{\Phi}(t) (1+it) \int_0^1 dq \times \medskip \\
e^{-(1-q)(1+it)H_\omega}[X,H_\omega] e^{-q(1+it)H_\omega},
\end{array}
\end{equation}
with $\tilde{\Phi}(t)=2\pi \int_{-\infty}^\infty dx e^{(1+it)x}\Phi(x)$ being the Laplace transform of $\Phi(x)=e^{-2\pi i F(x)}$-$1$, which is a smooth function with support in the spectral interval $s$. Then:
\begin{equation}
\begin{array}{c}
\mbox{Ind}\{\pi_+ U_\omega \pi_+ \} = \frac{1}{b_1}\int dP(\omega) \int dt \ \tilde{\Phi}(t) (1+it) \int_0^1 dq \times \medskip \\ 
\mbox{tr}_0\{(f(U_{\omega})-f(I))U_\omega 
e^{-(1-q)(1+it)H_\omega}[X,H_\omega] \pi_s e^{-q(1+it)H_\omega} \pi_s\}. 
\end{array}
\end{equation}
Since $\pi_s \chi(X)$ is Hilbert-Schmidt for any smooth $\chi$ with compact support, we can use Property 2 to move $e^{-q(1+it)H_\omega}\pi_s$ all the way to the left, inside $\mbox{tr}_0$. Since all the operators to the left of $[X,H_\omega]$ commute, we obtain
 \begin{equation}
 \begin{array}{c}
\mbox{Ind}\{\pi_+ U_\omega \pi_+\} = \frac{1}{b_1}\int dP(\omega)  \int dt \ \tilde{\Phi}(t) (1+it) \times \medskip \\ 
\mbox{tr}_0\{(f(U_{\omega})-f(I))U_\omega 
 e^{-(1+it)H_\omega}[X,H_\omega] \pi_s \} \medskip \\
= \frac{2 \pi i}{b_1}\int dP(\omega)  
\mbox{tr}_0\{(f(U_{\omega})-f(I))U_\omega F'(H_\omega)[X,H_\omega] \pi_s \} \nonumber 
\end{array}
\end{equation}
We now take
\begin{equation}
f(z)=\frac{z-1}{z-1+\epsilon}\frac{1}{z}, \ \epsilon>0,
\end{equation}
for which $b_1$=1. Then
\begin{equation}
\frac{1}{b_1}(f(U_\omega)-f(I))U_\omega=(U_\omega-I)(U_\omega-(1-\epsilon)I)^{-1},
\end{equation}
and by taking the limit $\epsilon \rightarrow 0$, we obtain:
\begin{equation}
\mbox{Ind}\{\pi_+ U_\omega \pi_+\} = 2 \pi i\int dP(\omega)  
\mbox{tr}_0\{F'(H_\omega)[X,H_\omega] \pi_s \}.
\end{equation}
This is precisely the trace of the theorem, since $\pi_s$ is equal to the identity on the spectral space corresponding to the support of $F'(x)$. \medskip

\noindent{\it Proof of the Non-Commutative Residue Theorem.} In the following, we use the notation $\nabla U_\omega$ for $[X,U_\omega]$. Let us take first $f(z)=z^k$. Then
\begin{equation}
[U_\omega^k-I]\nabla U_\omega=\nabla[(k+1)^{-1}(U_\omega^{k+1}-I)-(U_\omega-I)]
\end{equation}
plus terms that vanish when the operation $\int dP(\omega) \mbox{tr}_0$ is considered. Indeed,
\begin{equation}
\begin{array}{c}
[U_\omega^k-I]\nabla U_\omega-\nabla[(k+1)^{-1}(U_\omega^{k+1}-I)-(U_\omega-I)] \medskip \\
=-\frac{1}{k+1}U_\omega^k\sum_{m=0}^k [U_\omega^{-m}\nabla U_\omega (U_\omega^m -I)+ (U_\omega^{-m}-I) \nabla U_\omega].
\end{array}
\end{equation}
Now observe that since $\chi(X)\nabla U_\omega$ is Hilbert-Schmidt so is $\chi(X)U_\omega^{k-m} \nabla U_\omega$ and for this reason, after considering the operation $\int dP(\omega) \ \mbox{tr}_0 [.]$, we can apply Property 2 to move $(U_\omega^m -I)$ of the first term in the second row to the left of $\nabla U_\omega$. This way, the first term in the second row becomes minus the second one and the entire expression vanishes when we take the operation $\int dP(\omega)  \mbox{tr}_0 [.]$. Furthermore, since 
\begin{equation}
(k+1)^{-1}(U_\omega^{k+1}-I)-(U_\omega-I)=(U_\omega-1)g(U_\omega)(U_\omega-1),
\end{equation}
with $g(U_\omega)$ bounded, it follows that 
\begin{equation}
\chi(X)[(k+1)^{-1}(U_\omega^{k+1}-I)-(U_\omega-I)]\chi(X)
\end{equation}
is trace class with uniformly bounded trace class norm and from Property 3 we can conclude that:
\begin{equation}
\int dP(\omega) \mbox{tr}_0\{[U_\omega^k-I]\nabla U_\omega\}=0.
\end{equation}
Now we take $f(z)=z^{-k}$, $k>2$. In this case:
\begin{equation}
[U_\omega^{-k}-I]\nabla U_\omega=\nabla[(-k+1)^{-1}(U_\omega^{-k+1}-I)-(U_\omega-I)]
\end{equation}
plus terms that vanish when the operation $\int dP(\omega) \mbox{tr}_0[.]$ is considered. For $k>2$, we have
\begin{equation}
(-k+1)^{-1}(U_\omega^{-k+1}-I)-(U_\omega-I)=(U_\omega-1)g'(U_\omega)(U_\omega-1),
\end{equation}
with $g'(U_\omega)$ bounded, so from Property 3
\begin{equation}
\int dP(\omega) \mbox{tr}_0\{[U_\omega^{-k}-I]\nabla U_\omega\}=0.
\end{equation}
We consider now a general $f(z)$ and using its Laurent expansion, we write:
\begin{equation}
f(z)=\sum\limits_{n=1,M} b_n z^{-n} + \sum\limits_{n=0,M} a_n z^n+R_{-}^M(z)+R_{+}^M(z),
\end{equation}
with
\begin{equation}
R_{-}^M(z)=\sum\limits_{n>M} b_n z^{-n}, \ \ R_{+}^M(z)=\sum\limits_{n>M} a_n z^n
\end{equation}
Then we have:
\begin{equation}
\begin{array}{c}
[f(U_\omega)-f(I)]\nabla U_\omega=\sum\limits_{n=1}^M [b_n [U_\omega^{-n}-I]\nabla U + a_n[U_\omega^n-I]\nabla U_\omega]   \medskip \\
+[R_{-}^M(U_\omega)-R_{-}^M(I)]\nabla U_\omega 
+[R_{+}^M(U_\omega)-R_{+}^M(I)]\nabla U_\omega.
\end{array}
\end{equation}
Using the previous results, we can conclude at this step that:
\begin{equation}
\begin{array}{c}
\int dP(\omega)\mbox{tr}_0\{[f(U_\omega)-f(I)]\nabla U_\omega)\}-b_1 \int dP(\omega)\mbox{tr}_0\{[U_\omega^{-1}-I]\nabla U_\omega\} \medskip \\
=\int dP(\omega)\mbox{tr}_0\{([R_{-}^M(U_\omega)-R_{-}^M(I)]+[R_{+}^M(U_\omega)-R_{+}^M(I)])\nabla U_\omega\}.
\end{array}
\end{equation}
We continue as follows:
\begin{equation}
\begin{array}{c}
\int dP(\omega)\mbox{tr}_0\{([R_{-}^M(U_\omega)-R_{-}^M(I)]+[R_{+}^M(U_\omega)-R_{+}^M(I)])\nabla U_\omega\} \medskip \\
=\int dP(\omega)\mbox{Tr}\{\chi(X)(U_\omega^{-1}-1)(A+B) \nabla U_\omega\chi(X)\},
\end{array}
\end{equation}
with 
\begin{eqnarray}
A=-[R_{-}^M(U_\omega)-R_{-}^M(I)]/[U_\omega^{-1}-I], \  \|A\| \leq \sum\limits_{n>M} n b_n
\end{eqnarray}
and
\begin{eqnarray}
B=[R_{+}^M(U_\omega)-R_{+}^M(I)]/[U_\omega-I), \ \|B\| \leq  \sum\limits_{n>M} n a_n.
\end{eqnarray}
The conclusion is that:
\begin{equation}\label{upbound1}
\begin{array}{c}
|\int dP(\omega)\mbox{tr}_0\{[f(U_\omega)-f(I)]\nabla U_\omega)\}-b_1 \int dP(\omega)\mbox{tr}_0\{[U_\omega^{-1}-I]\nabla U_\omega\} | \medskip \\ 
\leq \int dP(\omega) \|\chi(X)(U_\omega-1)\|_{HS} \|\nabla U_\omega \chi(X)\|_{HS} \sum\limits_{n=M+1}^\infty (n b_n+n a_n),
\end{array}
\end{equation}
and the right side above goes to zero as $M\rightarrow \infty$ since $f(z)$ is analytic in a strip around the unit circle.

\section{Applications}

As an application we consider the quantization of differential conductance in atomic wires, which now is a well established experimental fact \cite{Terabe:2005cr}. We define the transport problem by following Ref.~\cite{Alekseev:1998p600}, where the quantization was related to an anomalous commutator whose trace fails to vanish. Here, we go one step further and connect the quantization to the index of a certain operator.

We consider first the 1D free, spinless electron gas, described by the Hamiltonian $H_0$=$-\frac{\hbar^2}{2m}\frac{d^2}{dx^2}$. For this we consider a pair of charge operators $Q_+$ and $Q_-$ such that:
\begin{equation}
[H_0,Q_\pm]=0 \ , \ \ [Q_+,Q_-]=0.
\end{equation}
$Q_\pm$ are taken as the spectral projectors of the linear momentum $p=\frac{\hbar}{i}\frac{d}{dx}$ onto its positive/negative parts of the spectrum.
The driving Hamiltonian at finite electric bias potential $v$ is
\begin{equation}\label{driving}
H_v = -\frac{\hbar^2}{2m}\frac{d^2}{dx^2}-\frac{ev}{2}(Q_+-Q_-).
\end{equation}
The observable $X$ is taken as the position operator $x$ and the unitary group $u_a$ is the translation by $a$. $H_v$ is invariant to translations, so the set $\Omega$ reduces to just one point and the average over $\omega$ becomes irrelevant. The electric current is
\begin{equation}
e\left \langle \frac{dx}{dt}\right \rangle_v = -\frac{ie}{\hbar}  \ \mbox{tr}_0 \{\Phi(H_v)[x,H_v]\},
\end{equation}
where $\Phi(H_v)$ is some, possibly non-equilibrium, statistical distribution. The only requirements we have on $\Phi$ is that it is a smooth function equal to 1 below an energy $E_m>\frac{ev}{2}$ and equal to 0 above an arbitrary large but finite energy $E_M>E_m$ (see Fig.~3). The differential conductance is given by
\begin{equation}\label{dcond}
\begin{array}{c}
g=e\frac{d}{dv}\left \langle \frac{dx}{dt}\right \rangle_v \medskip \\
= -\frac{ie^2}{2\hbar}  \ \mbox{tr}_0 \{\Phi'(H_v)\Delta Q[x,H_v]+\Phi(H_v)[x,\Delta Q]\},
\end{array}
\end{equation}
where $\Delta Q = Q_+-Q_-$.\medskip

\begin{figure}
 \center
 \includegraphics[width=7cm]{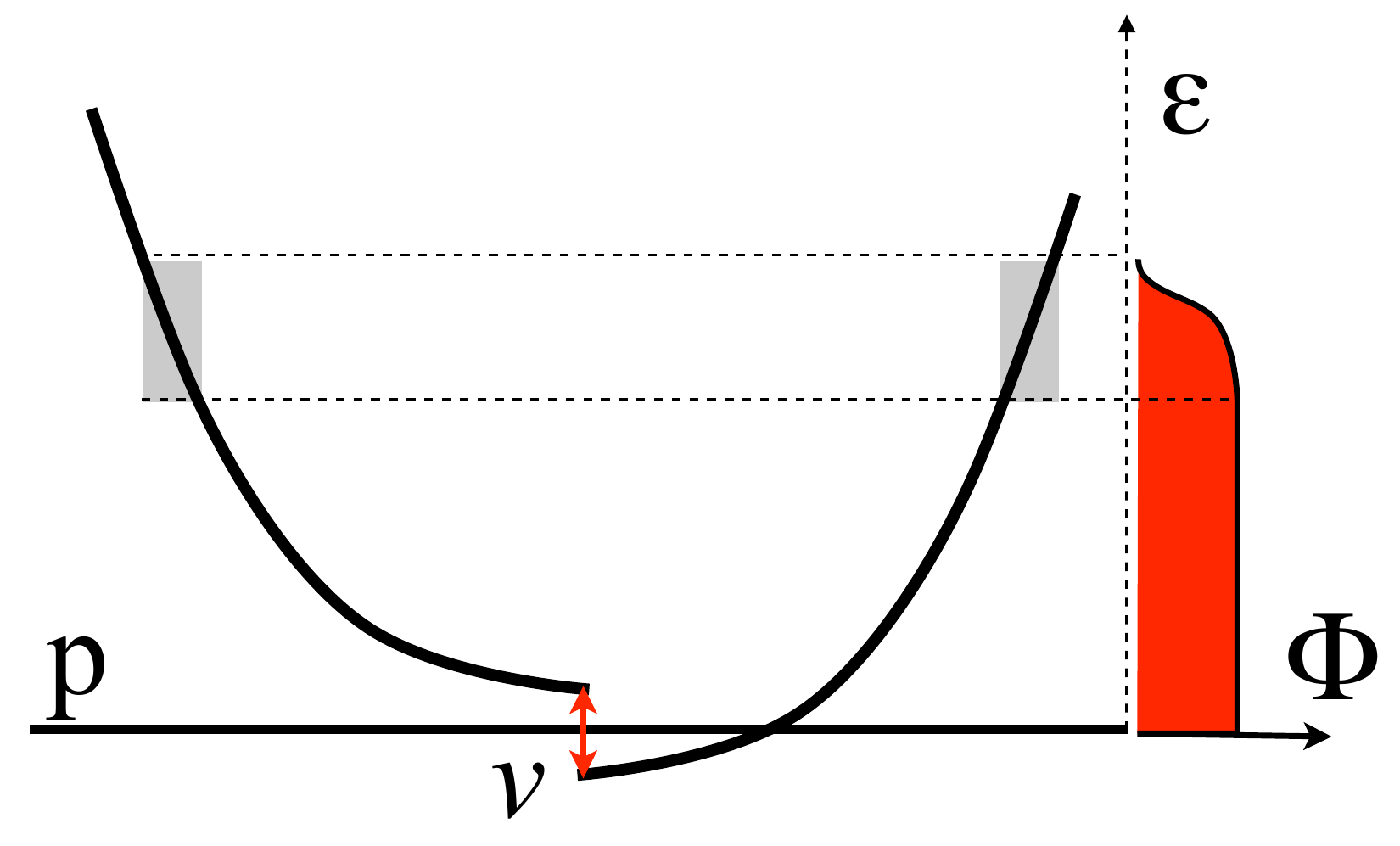}
 \caption{A plot of the energy spectrum of $H_v$ as function of $p$: $\epsilon=p^2/2m +(ev/2)  \mbox{sign}(p)$. The figure also shows an admissible distribution $\Phi$. Given the particular form of $\Phi$, the function $e^{-2\pi i \Phi(p^2/2m -ev\mbox{\tiny{sign}}(p)/2)\mbox{\tiny{sign}}(p)} -1$, appearing in the proof of Proposition 3, is nonzero only in the shaded regions.}
 \label{division}
\end{figure}

\noindent{\bf Proposition 3.}  Define $U=\exp(-2\pi i \Phi(H_v)\Delta Q)$. Then:
\begin{equation}
g= -\frac{e^2}{2h}\mbox{Ind}\{\pi_+ U \pi_+ \}.
\end{equation}
\noindent {\it Proof.} We can verify directly that $(U - I)\pi_\Delta(x_0)$ is Hilbert-Schmidt. Indeed, we have:
\begin{equation}
\begin{array}{c}
\pi_\Delta(x_0) (U^*-I)(U - I)\pi_\Delta(x_0) \medskip \\ 
=2 \pi_\Delta(x_0) [I-\cos(2\pi \Phi(H_v)\Delta Q)] \pi_\Delta(x_0),
\end{array}
\end{equation}
and the trace of the latest is
\begin{eqnarray}
2 \Delta \int dp  \left [1-\cos\left(2\pi \Phi\left(\frac{p^2}{2m}-\frac{ev}{2}\mbox{sign}(p)\right )\mbox{sign}(p)\right)\right ],
\end{eqnarray}
which is finite since the integrand is non-zero only on a finite set. We can also compute the kernel $K_\Delta (x,x')$ explicitly:
\begin{equation}
\begin{array}{c}
K_\Delta (x,x')=\frac{\hbar}{2\pi}\int\limits _{x'}^{x'+\Delta}d\xi'\int\limits_x^{x+\Delta}d\xi \ \times\medskip \\
\left|\int\limits_{-\infty}^{\infty} dp \left ( e^{-2\pi i \Phi(p^2/2m -ev\mbox{\tiny{sign}}(p)/2)\mbox{\tiny{sign}}(p)} -1 \right )e^{\frac{i}{\hbar}p(\xi-\xi')} \right |^2.
\end{array}
\end{equation}
Because of our assumptions on $\Phi(x)$, $e^{-2\pi i \Phi(p^2/2m -ev\mbox{\tiny{sign}}(p)/2)\mbox{\tiny{sign}}(p)} -1$  has support in the two intervals shaded in Fig. 3, which are away from the singularity at $p$=0 and consequently this function is smooth.  Thus the integral of the last row is rapidly decaying with the separation $|\xi-\xi'|$. More precisely, we can choose arbitrarily large $\alpha$ in ${\cal P}$. It is important to notice that second condition in ${\cal P}$ will be violated if the support of $\mbox{d}\Phi/\mbox{d}t$ would have overlap with the two spectral edges at the bottom of the spectrum of $H_v$. The integral of the last row is also a smooth function of $\xi$ and $\xi'$, thus we can choose $\beta$=2 in ${\cal P}$. Then the proof given above applies all the way to Eq.~\ref{ref1}, 
\begin{equation}
\begin{array}{c}
 \mbox{Ind}\{\pi _+U \pi_+\} =-\frac{1}{b_1}  \mbox{tr}_0\{ (f(U)-f(I))[x, U]\pi_s^3 \}. 
\end{array}
\end{equation}
Here we inserted $\pi_s^3$ instead of $\pi_s^2$. We can write $U=e^{-2\pi i A}$ with $A=\Phi(H_v)\Delta Q$ and treat the commutator via the Duhamel's identity. Following the same steps as in the main proof, we then obtain:
\begin{equation}
\begin{array}{c}
\mbox{Ind}\{\pi_+ U \pi_+\}= \frac{2 \pi i}{b_1}  
\mbox{tr}_0\{(f(U)-f(I))U [x, A] \pi_s^2\} \medskip \\
  = \frac{2 \pi i}{b_1}  
\mbox{tr}_0\{(f(U)-f(I))U (\Phi(H_{v})[x, \Delta Q]+[x, \Phi(H_{v})]\Delta Q) \pi_s^2\}.
\end{array}
\end{equation}
We repeat the procedure for the commutator $[x, \Phi(H_{v})]$ to obtain:
\begin{equation}
\begin{array}{c}
\mbox{Ind}\{\pi_+ U \pi_+\}=\frac{ 2 \pi i}{b_1}  
\mbox{tr}_0\{(f(U)-f(I))U \Phi(H_{v})[x, \Delta Q] \pi_s^2\} \medskip \\
  +\frac{ 2 \pi i}{b_1}  
\mbox{tr}_0\{(f(U)-f(I))U \Phi'(H_v)\Delta Q[x, H_{v}] \pi_s\}.
\end{array}
\end{equation}
We can complete the proof by taking $f(z)$ like in the main proof.\medskip

\begin{figure}
 \center
 \includegraphics[width=7cm]{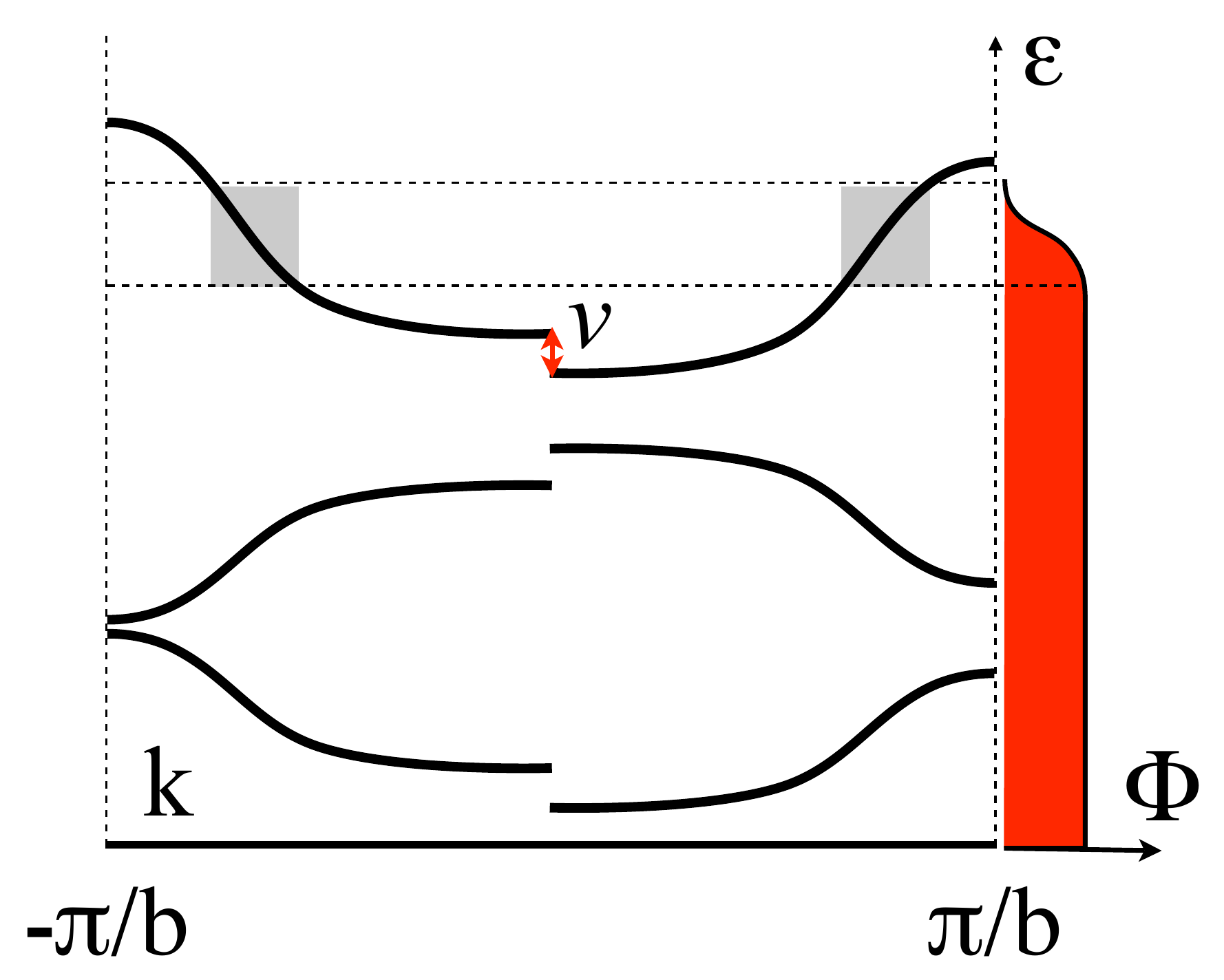}
 \caption{A plot of the energy spectrum of $H_{\omega,v}$ as function of $k$: $\epsilon=\epsilon_k +(ev/2)  \mbox{sign}(k)$. The figure also shows an admissible distribution $\Phi$. Given the particular form of $\Phi$, the function $e^{\epsilon_k +(ev/2)  \mbox{\tiny{sign}}(k))\mbox{\tiny{sign}}(p)} -1$ is nonzero only in the shaded region.}
\end{figure}

We now state a similar result for the non-interacting electron gas in a periodic potential. The zero bias Hamiltonian is:
\begin{equation}
H_\omega= -\frac{\hbar^2}{2m}\frac{d^2}{dx^2}+V(R_x\omega),
\end{equation}
where $V$ is a smooth function defined on a circle of length $b$, $\omega$ is a point on this circle and $R_x$ represents the rotation of the circle by an angle $x$. We define the charge operators using the Bloch fibration, i.e. the unitary transformation 
\begin{equation}
U: L^2({\bf R})\rightarrow \oplus_{k\in [-\frac{\pi}{b},\frac{\pi}{b}]} L^2([0,b])
\end{equation}
such that
\begin{equation}
UH_\omega U^* = \oplus_{k\in [-\frac{\pi}{b},\frac{\pi}{b}]} H_\omega(k),
\end{equation}
with $H_\omega(k)$ the usual Bloch Hamiltonians. Let $P_{\omega,n}(k)$ be the  spectral projectors onto the eigenvalues $\epsilon_{n,k}$ of the Bloch Hamiltonians. The charge operators can be defined as:
\begin{eqnarray}
Q_{\omega+} = U^*\left [\oplus_{\mbox{\tiny{d}}\epsilon_k / \mbox{\tiny{d}}k>0} \sum_n \ P_{\omega,n}(k)\right ]U, \\
Q_{\omega-} = U^* \left [\oplus _{\mbox{\tiny{d}}\epsilon_k / \mbox{\tiny{d}}k<0}\sum_n  \ P_{\omega,n}(k)\right]U.\nonumber 
\end{eqnarray}
These charge operators simply split the Hilbert space into right and left moving Bloch waves. We define the driving Hamiltonian, $H_{\omega,v}$ at finite electric bias potential $v$ as above. 

We consider now a smooth statistical distribution $\Phi(x)$ with the constraint that we now explain. At finite bias potential, the right/left moving states in each band move up/down by $ev/2$. Thus, at finite bias potential, 2 spectral edges are generated for each band as illustrated in Fig.~4. The constrained we need to impose on $\Phi(t)$ is that the support of $\mbox{d}\Phi/\mbox{d}t$ does not contain any of the spectral edges of $H_{\omega,v}$. This constraint ensures that $e^{-2\pi i \Phi(\epsilon_k +ev\mbox{\tiny{sign}}(k)/2)\mbox{\tiny{sign}}(k)}-1$ has a support that is away from the singularities at $k$=0 (see the shaded regions in Fig.~4). With the charge current and the differential conductance $g(\omega)$ defined as before, we have:\medskip 

\noindent{\bf Proposition 4.}  Define $U_\omega=\exp(-2\pi i \Phi(H_{\omega,v})\Delta Q_\omega)$. Then:
\begin{equation}
\frac{1}{b}\int d\omega \ g(\omega)= -\frac{e^2}{2h}\mbox{Ind}\{\pi_+ U_\omega \pi_+ \}. \medskip
\end{equation}

\noindent A similar result can be formulated for periodic molecular chains in 3 dimensions. The index in Proposition 3 and 4 is 2.

\section{Discussion}

We want to comment on the properties ${\cal P}$. In Ref.~\cite{Prodan:2008ai} we give a fairly general methodology for proving estimates of the form Eq.~\ref{decay}. The estimates were derived for general tight binding (discrete) Hamiltonians with a clean bulk insulating gap. What we have learned from this application is that, if certain exponential decay estimates on the bulk Hamiltonian can be derived, then the estimate of Eq.~\ref{decay} can be obtained via fairly standard techniques. For continuous Schrodinger operators that include magnetic fields and scalar potentials and have a gap in the spectrum, exponential decay estimates have been derived in Ref.~\cite{Prodan:2006cr}. It appears to us that we can use these decay estimates and repeat the steps of Ref.~\cite{Prodan:2008ai} to prove ${\cal P}$ for half-plane continuous magnetic Schrodinger operators with weak random potential.

The result of our main theorem can be an effective tool for characterization of the edge states, which was also part of our motivation for looking into this problem. The problem of edge states received a renewed attention \cite{Qi:2006kx}, since the discovery of the Spin-Hall effect \cite{Hirsch:1999ve, Kane:2005vn, Kane:2005ys, B.A.-Bernevig:2006zr}. For the edge states problem in insulators, one will take the spectral interval $s$ as the bulk gap and $X$ as an appropriate observable (in the Hall problem that will be the coordinate along the edge). To see how the formalism works, notice that the right hand side of  Eq.~\ref{main} is independent of the shape of $F(x)$. Thus, if there is a spectral gap inside the interval $s$, we can choose $F(x)$ such that the operator $U_\omega$-$I$  is identically 0 and, consequently, the right hand side of Eq.~\ref{main} is zero. Thus, if we can show that the index is different from zero, that will automatically imply that the interval $s$ is filled with (edge) states. To compute the index, one can continuously deform the system until it becomes either analytically or numerically solvable. To give an example, for a periodic system with a rough edge, one will try a deformation into a system with a smooth, periodic edge, in which case the spectrum and the current can be obtained from a standard band structure calculation. Of course, one has to show that ${\cal P}$ holds true during the deformation (which is not difficult). What is new in this argument is that the index can be computed even if the gap closes and remain closed during the deformation. This scenario was excluded in Ref.~\cite{Qi:2006kx} and it seems that it is not manageable by the existing techniques.  Such scenarios can happen very often due to shifts of surface states during the deformation, states that may or may not have anything to do with the quantization. 

\medskip

{\bf References}

\bibliographystyle{iopart-num}


\end{document}